\documentclass[aps,prl,twocolumn,showpacs,superscriptaddress,floatfix]{revtex4-1}
\pdfoutput=1

\usepackage{amsmath,amssymb}
\usepackage{graphicx}
\usepackage{color}
\usepackage{rotating}
\usepackage{bm}
\usepackage{setspace}
\usepackage{hyperref}

\hypersetup{
colorlinks=true,
urlcolor= blue,
citecolor=blue,
linkcolor= blue,
bookmarks=true,
bookmarksopen=false,
}

\begin{document}

%\preprint{}

\title{Dumbbell Defects in FeSe Films:\\ A Scanning Tunneling Microscopy and First-Principles Investigation}
\author{Dennis Huang}
\affiliation{Department of Physics, Harvard University, Cambridge, Massachusetts 02138, USA}
\author{Tatiana A. Webb}
\affiliation{Department of Physics, Harvard University, Cambridge, Massachusetts 02138, USA}
\affiliation{Department of Physics \& Astronomy, University of British Columbia, Vancouver, British Columbia V6T 1Z1, Canada}
\author{Can-Li Song}
\altaffiliation[]{Present address: State Key Laboratory of Low-Dimensional Quantum Physics, Department of Physics, Tsinghua University, Beijing 100084, China}
\affiliation{Department of Physics, Harvard University, Cambridge, Massachusetts 02138, USA}
\author{Cui-Zu Chang}
\affiliation{Francis Bitter Magnet Laboratory, Massachusetts Institute of Technology, Cambridge, Massachusetts 02139, USA}
\author{Jagadeesh S. Moodera}
\affiliation{Francis Bitter Magnet Laboratory, Massachusetts Institute of Technology, Cambridge, Massachusetts 02139, USA}
\affiliation{Department of Physics, Massachusetts Institute of Technology, Cambridge, Massachusetts 02139, USA}
\author{Efthimios Kaxiras}
\email[]{kaxiras@physics.harvard.edu}
\affiliation{Department of Physics, Harvard University, Cambridge, Massachusetts 02138, USA}
\affiliation{John A. Paulson School of Engineering and Applied Sciences, Harvard University, Cambridge, Massachusetts 02138, USA}
\author{Jennifer E. Hoffman}
\email[]{jhoffman@physics.harvard.edu}
\affiliation{Department of Physics, Harvard University, Cambridge, Massachusetts 02138, USA}
\affiliation{Department of Physics \& Astronomy, University of British Columbia, Vancouver, British Columbia V6T 1Z1, Canada}

\date{\today}


\begin{abstract}
The properties of iron-based superconductors (Fe-SCs) can be varied dramatically with the introduction of dopants and atomic defects. As a pressing example, FeSe, parent phase of the highest-$T_c$ Fe-SC, exhibits prevalent defects with atomic-scale ``dumbbell'' signatures as imaged by scanning tunneling microscopy (STM). These defects spoil superconductivity when their concentration exceeds 2.5\%. Resolving their chemical identity is prerequisite to applications such as nanoscale patterning of superconducting/nonsuperconducting regions in FeSe, as well as fundamental questions such as the mechanism of superconductivity and the path by which the defects destroy it. We use STM and density functional theory to characterize and identify the dumbbell defects. In contrast to previous speculations about Se adsorbates or substitutions, we find that an Fe-site vacancy is the most energetically favorable defect in Se-rich conditions, and reproduces our observed STM signature. Our calculations shed light more generally on the nature of Se capping, the removal of Fe vacancies via annealing, and their ordering into a $\sqrt{5}$$\times$$\sqrt{5}$ superstructure in FeSe and related alkali-doped compounds.
\end{abstract}

\pacs{}

\maketitle

%I. Introduction (1): FeSe
FeSe, a member of the iron-based superconductors (Fe-SCs) with the simplest stoichiometry, lies at the vanguard of high-$T_c$ materials. On one hand, its anomalous parent phase, with no static magnetic order~\cite{Medvedev_NatMat_2009}, poses a fresh theoretical challenge~\cite{Glasbrenner_NatPhys_2015, Wang_NatPhys_2015, Chubukov_PRB_2015, Yu_PRL_2015}. On the other hand, its plain, 2D-layered structure lends itself to bottom-up, nanoscale engineering of its electronic properties. As a striking example, monolayer FeSe interfaced with SrTiO$_3$~\cite{Wang_CPL_2012} exhibits an order-of-magnitude enhancement in its transition temperature $T_c$ (up to 109 K~\cite{Ge_NatMat_2014}) compared to its bulk value (8 K~\cite{Hsu_PNAS_2008}). Similar $T_c$ boosts up to 48 K have also been attained by depositing K adatoms~\cite{Miyata_NatMat_2015, Wen_NatComm_2016}, opening the door to all kinds of adatom modifications of FeSe. 

%II. Introduction (2): Defect engineering 
More generally, defects in Fe-SCs are crucial to control $T_c$~\cite{Kamihara_JACS_2008, Yeh_EPL_2008}, raise the critical current $J_c$ through vortex pinning~\cite{Yin_PRL_2009, Song_PRB_2013}, and also serve as microscopic probes of pairing symmetry~\cite{Balatsky_RMP_2006, Kashiwaya_RPP_2000}. Furthermore, defect effects are typically enhanced in 2D systems. An ultimate goal is to control precise placement of atomic defects, possibly through scanning probe lithography, as has been achieved with hydrogenated graphene~\cite{Sessi_NL_2009}, P dopants in Si~\cite{Weber_Science_2012}, and Mn dopants in GaAs~\cite{Kitchen_Nat_2006}. To similarly pattern nanostructures in FeSe, an atomistic understanding of defect formation in this material is needed. 

\begin{figure*}[t]
\includegraphics[scale=1]{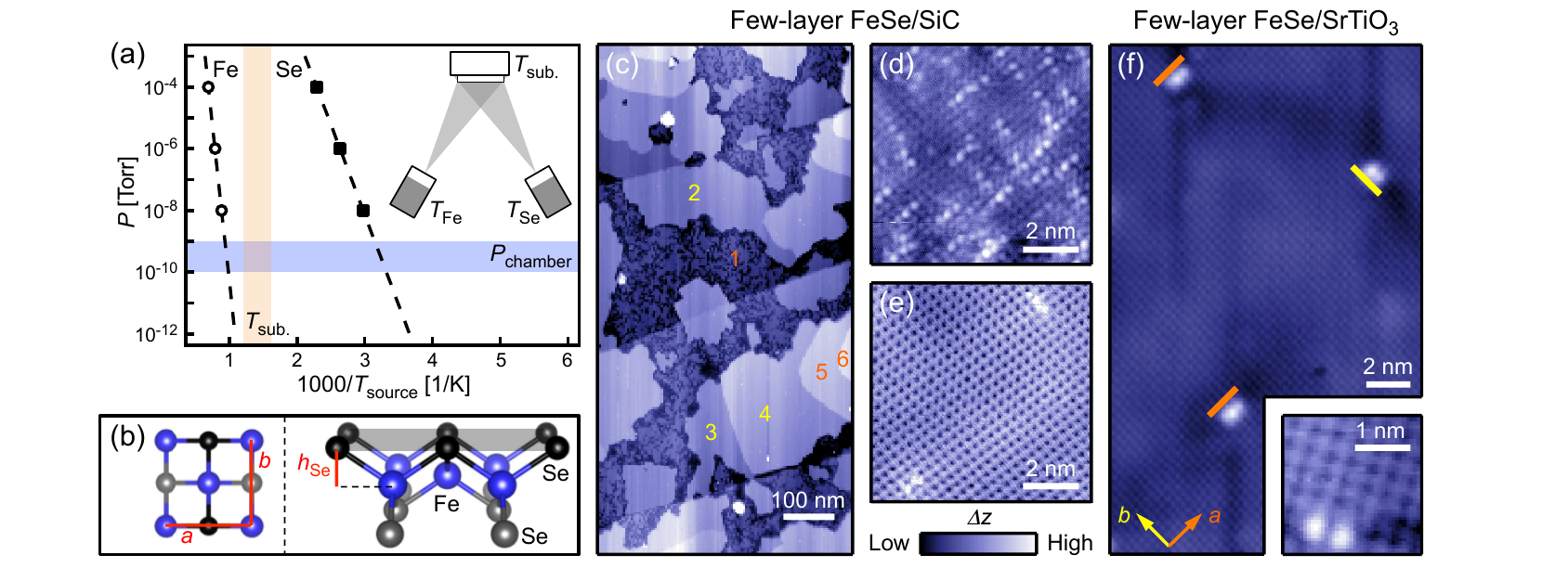}
\caption{(a) Sublimation curves for elemental Fe and Se, reproduced from Ref.~\cite{Oxford_2008}. Shaded horizontal and vertical bars mark typical chamber pressures ($10^{-10}$$-$$10^{-9}$ T) and typical substrate temperatures (350$-$550 $^{\circ}$C). The inset schematic illustrates FeSe growth via molecular beam epitaxy. (b) Crystal structure of a single layer of FeSe, viewed from the top and side. The shaded plane marks top-layer Se atoms imaged by scanning tunneling microscopy. (c)-(e) Topographic images of few-layer FeSe/SiC. (c) FeSe exhibits island growth on SiC. Numbers indicate unit cell thicknesses. Set point: 4 V, 5 pA; $T$ = 79 K. (d) Dumbbell defects in few-layer FeSe/SiC. Set point: 10 mV, 100 pA; $T$ = 84 K. (e) Same film as in (d), but after annealing at $\sim$450 $^{\circ}$C for 2.5 h. Set point: 10 mV, 5 pA; $T$ = 83 K. (f) Dumbbell defects in few-layer FeSe/SrTiO$_3$. Orange and yellow bars mark two possible orientations of the dumbbells. Set point: 100 mV, 5 pA; $T$ = 6.8 K. For inset: 100 mV, 5 pA; $T$ = 6.2 K.}
\label{Fig1}
\end{figure*}

%III. Introduction (3): Dumbbell defects
As an intriguing and urgent example, FeSe films grown by molecular beam epitaxy (MBE) exhibit prevalent defects with atomic-scale ``dumbbell'' signatures as imaged by scanning tunneling microscopy (STM) (also called geometric dimers in Ref.~\cite{Song_PRL_2012}). They consist of two bright lobes on adjacent top-layer Se sites [Figs.~\ref{Fig1}(d)-(f)]. Their concentration is highly tunable, increasing with excess Se flux and decreasing with substrate temperature. Importantly, superconductivity emerges only when their concentration falls below 2.5\%~\cite{Song_PRB_2011}. Despite the structural simplicity of FeSe, it is still unknown whether these dumbbell defects are Se adsorbates, antisites, interstitials, or some other type of defect. Their identity is crucial to determine whether or not they can be engineered to define superconducting/nonsuperconducting regions in FeSe for nanoscale applications.

%IV. Overview
Here we present an STM characterization of dumbbell defects and an exhaustive, first-principles investigation of candidate defect configurations. Using density functional theory (DFT), we find that Fe vacancies have the lowest formation energy. Furthermore, our modeling shows that they perturb orbitals on neighboring Se sites, producing dumbbell signatures when imaged by STM. Based on nudged elastic band calculations and 2D random walk simulations, we explain how Fe vacancies can diffuse to the edge of terraces during vacuum annealing, consistent with experimental observations of reduced dumbbell density after annealing. We further discuss implications for Se capping of FeSe films for \textit{ex-situ} applications. We also connect our results to previous questions of vacancy ordering in FeSe and related alkali-doped compounds.

%V. MBE Growth [Fig. 1]
\textbf{Methods.} Films of FeSe were deposited via MBE on 6H-SiC(0001) and SrTiO$_3$(001) substrates, following established recipes~\cite{Song_Science_2011, Song_PRB_2011, Wang_CPL_2012, Huang_PRL_2015}. The greater volatility of one element (Se) over the other (Fe) motivates two conditions for stoichiometric growth~\cite{Song_PRB_2011}: First, by setting the substrate temperature between the source temperatures, $T_{\textrm{Fe}} > T_{\textrm{substrate}} > T_{\textrm{Se}}$, impinging Fe with temperature $\sim T_{\textrm{Fe}}$ will be adsorbed with sticking coefficient close to unity, while impinging Se can stick only if they bind to free Fe on the substrate [Fig.~\ref{Fig1}(a)]. Second, to compensate for high Se losses and to mitigate excess Fe clustering, typical molar flux ratios $\Phi_{\textrm{Se}}/\Phi_{\textrm{Fe}}$ range from 5 to 20. 

%VI. STM characterization of dumbbell defects [Fig. 1]
Post growth, the films were transfered \textit{in situ} to a homebuilt STM and imaged at liquid nitrogen/helium temperatures. From Figs.~\ref{Fig1}(d)-(f), we enumerate several characteristics of the dumbbell defects: First, their prevalence over any other kinds of defects suggests they are energetically favorable. In few-layer FeSe, this observation is independent of substrate, SiC [Figs.~\ref{Fig1}(c)-(e)] or SrTiO$_3$ [Fig.~\ref{Fig1}(f)]. (We note that single-layer FeSe/SrTiO$_3$, with vastly different superconducting properties, exhibits a different set of defects~\cite{Huang_PRL_2015, Huang_PRB_2016}.) Similar dumbbell defects have also been imaged in FeSe crystals grown by vapor transport~\cite{Kasahara_PNAS_2014, Watashige_PRX_2015} and in Li$_{1-x}$Fe$_x$OHFeSe crystals grown by hydrothermal ion exchange~\cite{Du_NatComm_2016, Yan_arXiv_2015}. Second, the dumbbells are aligned along both the $a$ and $b$ axes of the 2-Fe unit cell [Figs.~\ref{Fig1}(d)-(f)], pointing to their independence from a structural orthorhombic distortion~\cite{McQueen_PRL_2009} and electronic nematic state in FeSe~\cite{Shimojima_PRB_2014, Nakayama_PRL_2014, Watson_PRB_2015, Zhang_PRB_2015} that break 90$^{\circ}$ rotational symmetry. Third, our STM measurements up to $T$ = 84 K with bias voltages 10$-$100 mV demonstrate that the dumbbell signatures persist well above the superconducting state. Fourth, the dumbbell defects can be removed upon annealing, leaving behind pristine FeSe [Fig.~\ref{Fig1}(e)]. 

\begin{table*}[t]
\center
\setlength{\tabcolsep}{10pt}
\begin{tabular}{ccccc}
\hline
 & Monolayer & Bilayer & Film & Bulk \\
 & & & (expt.) & (expt.) \\
\hline\hline
Functional: & GGA & GGA/DFT-D2 & \\
Supercell size: & 4$\times$4 & 3$\times$3 & \\
BZ sampling: & 2$\times$2$\times$1 & 4$\times$4$\times$1 & \\
$a$=$b$ [\AA]: & 3.69  & 3.64 & 3.8-3.9 & 3.7707\\
$c$ [\AA]: & & 5.47 & 5.5 & 5.521\\
$h_{\textrm{Se}}$ [\AA]: & 1.38 & 1.40 & & 1.472 \\
$c_{\textrm{supercell}}$ [\AA]: & 20 & 25 & &\\
\hline
\end{tabular}
\caption{Relaxed parameters of monolayer and bilayer FeSe supercells used to simulate defect configurations. $a$, $b$, $c$ are the crystal lattice constants for a 2-Fe unit cell, $h_{\textrm{Se}}$ is the internal Se height, and $c_{\textrm{supercell}}$ includes vacuum regions. Experimental values for films are based on STM. Experimental values for bulk crystals are based on X-ray powder diffraction~\cite{Bohmer_PRB_2013}.}
\label{Tab1}
\end{table*}

%VII. DFT Details [Table 1]
We performed DFT calculations using \texttt{VASP}~\cite{Kresse_CMS_1996, Kresse_PRB_1996}. We used the PBE exchange-correlation functional~\cite{Perdew_PRL_1996}, and the projector augmented wave (PAW) method, with Fe 4$s$, 3$d$ and Se 4$s$, 4$p$ electrons treated as valence. An energy cutoff of 450 eV and Methfessel-Paxton smearing~\cite{Methfessel_PRB_1989} with $\sigma$ = 0.1 eV were employed. We modeled defects within freestanding monolayer and bilayer FeSe supercells (details in Table~\ref{Tab1}), with full relaxation of internal atomic coordinates (corresponding to a magnitude of the force per atom $<$ 0.025 eV/\AA). To reproduce the experimental $c$-axis value, we included van der Waals corrections in the bilayer calculations using the DFT-D2 method~\cite{Grimme_JCC_2006}, with dispersion potential parameters taken from Ref.~\cite{Ricci_PRB_2013} (tested for bulk FeSe and FeTe). 

\begin{figure}[b]
\includegraphics[scale=1]{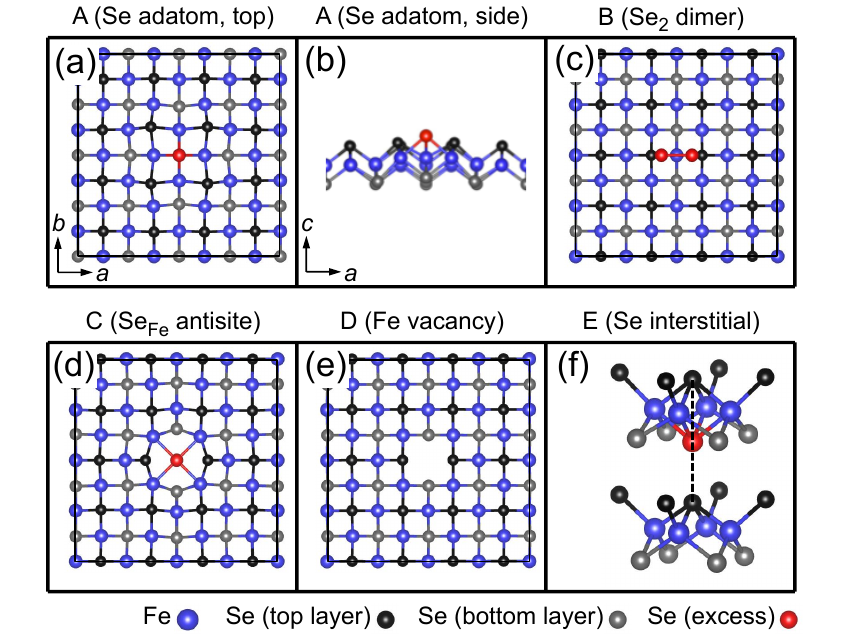}
\caption{Relaxed defect structures for monolayer FeSe 4$\times$4 supercells (a)-(e) . Solid-line boxes mark the supercell boundaries. For the interstitial configuration (f), only a fraction of the bilayer FeSe 3$\times$3 supercell is shown for clarity. Fe atoms are violet and top/bottom Se atoms are black/gray. Excess Se atoms are colored red for distinction.} 
\label{Fig2}
\end{figure}

%VIII. Se adatoms [Figs. 2(a), (b)]
\textbf{Results.} Given the correlation of dumbbell defects with excess Se flux, we examine candidate defects in which $N_{\textrm{Se}} > N_{\textrm{Fe}}$. Although the dumbbell signature is centered above an Fe site, we explore all possible binding sites for completeness. We begin with isolated Se adatoms as the simplest class of Se-rich defects. Among three adsorption sites (see Supporting Information), the hollow site in FeSe, directly above a bottom-layer Se atom, is most stable [Fig.~\ref{Fig2}(a)-(b)]. We compute the binding energy as 
\begin{equation}
\label{Eqadatom}
E_{\textrm{adatom}} = E(D) - E(0) - E_{\textrm{Se}},
\end{equation}
where $E(D)$ is the DFT total energy of the system including the adatom, $E(0)$ is the total energy of pristine FeSe within the same supercell, and $E_{\textrm{Se}}$ is the energy of an isolated Se atom. We find that $E_{\textrm{adatom}}$ = $-$3.14 eV ($-$3.02 eV) for monolayer (bilayer) FeSe, which suggests chemisorption. Examining the relaxed structure [Fig.~\ref{Fig2}(b)], we observe that the Se adatom comes within bonding distance of neighboring Fe atoms and induces local strain. Importantly, given that $T_c$ in Fe-SCs is highly sensitive to the Fe-Se/As height~\cite{Okabe_PRB_2010}, our result points to a possible microscopic explanation of why amorphous Se may be a poor capping material.

%IX. Se dimers [Fig. 2(c)]
We next examine adsorbed Se$_2$ dimers. We find that among five possible adsorption geometries (see Supporting Information), two are nearly degenerate, one of which has the Se$_2$ molecule centered above an Fe site [Fig.~\ref{Fig2}(c)]. Furthermore, with binding energy defined as 
\begin{equation}
E_{\textrm{dimer}} = E(D) - E(0) - E_{\textrm{Se}_2},
\end{equation}
where $E_{\textrm{Se}_2}$ is the energy of an isolated Se$_2$ molecule, we calculate $E_{\textrm{dimer}}$ = $-$0.39 eV ($-$0.69 eV) for monolayer (bilayer) FeSe. These values suggest that Se$_2$ dimers are weakly physisorbed and may have short adsorption lifetimes. We contrast this result to the case of GaAs(001)-(2$\times$4), where surface dangling bonds can stabilize adsorbed As$_2$ dimers or As$_4$ tetramers with calculated binding energies up to $-$1.6 eV~\cite{Morgan_PRL_1999}. Such dangling bonds are absent in the top layer of FeSe. As a side note, our DFT calculations suggest that adsorption can be enhanced if two surface dimers cluster into Se$_4$, but this would produce an unobserved four-lobe STM topographic signature.

\begin{figure}[t]
\includegraphics[scale=1]{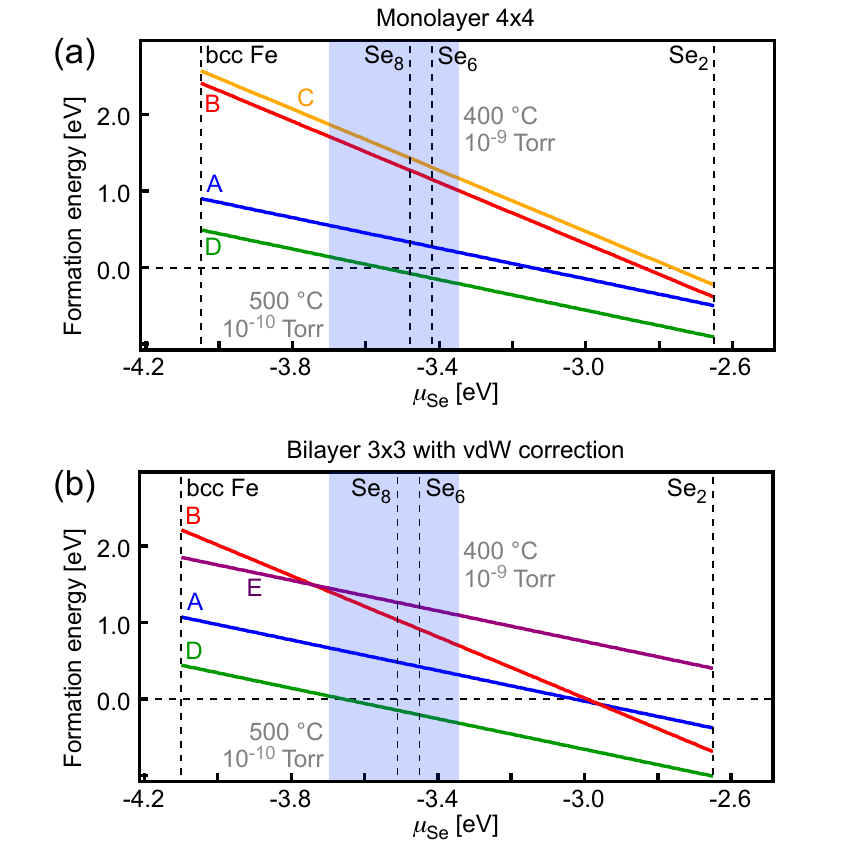}
\caption{Formation energies of defect configurations in the (a) 4$\times$4 supercell monolayer and (b) 3$\times$3 supercell bilayer FeSe. Capital letters correspond to labeled defects in Fig.~\ref{Fig2}. Assuming no condensation of bulk Fe (body-centered cubic) or Se (Se$_6$ and Se$_8$ rings), chemical potential values $\mu_{\textrm{Se}}$ are restricted between the dashed lines labeled bcc Fe and Se$_6$/Se$_8$. $\mu_{\textrm{Se}}$ is set to zero for an isolated Se atom. Alternatively, shaded blue regions mark estimated $\mu_{\textrm{Se}}$ values at typical substrate temperatures and Se partial pressures, using ideal gas approximations and tabulated thermodynamic quantities (see Supporting Information).}
\label{Fig3}
\end{figure}

%X. Antisite, vacancy, interstitial [Figs. 2(d)-(f)]
Alternatively, some studies have proposed that a perturbation at the Fe site (either an unknown repulsive potential~\cite{Choubey_PRB_2014} or Se subsitution~\cite{Li_JPCM_2014}) could affect the orbitals on neighbouring Se atoms and generate a dumbbell signature. We consider three possibilities: Se$_{\textrm{Fe}}$ antisites, Fe-site vacancies, and Se interstitials, perhaps binding to a surface-layer Fe atom from below. As seen in Fig.~\ref{Fig2}(d), the antisite configuration in the monolayer supercell produces pronounced distortions of nearby atoms. Fe atoms are pulled closer to the antisite, and Se atoms are pushed away. We note that the antisite could not be held in place in the bilayer supercell during structural relaxation. Figure~\ref{Fig2}(e) shows an Fe vacancy. Figure~\ref{Fig2}(f) shows the most stable Se interstitial configuration, where the excess Se atom lies beneath a top-layer Se site, not an Fe site.

%XI. Defect formation energies [Figs. 3(a), (b)]
To compare formation energies $E_f$ among the aforementioned defects with variable stoichiometry, we include the energetic costs of incorporating $n_{\textnormal{Fe}}$ ($n_{\textnormal{Se}}$) additional Fe (Se) atoms from a reservoir into the defect:
\begin{equation}\label{Eqformation}
E_f = E(D) - E(0) - n_{\textnormal{Fe}}\mu_{\textnormal{Fe}} - n_{\textnormal{Se}}\mu_{\textnormal{Se}}.
\end{equation}
Assuming quasi-equilibrium growth of FeSe and no bulk Fe or Se precipitation, we impose the following constraints on the chemical potentials: (i) $\mu_{\textnormal{Fe}} + \mu_{\textnormal{Se}} = \mu_{\textnormal{FeSe}}$; (ii) $\mu_{\textnormal{Fe}} < \mu_{\textnormal{Fe}}^{\textnormal{bulk}}$; (iii) $\mu_{\textnormal{Se}} < \mu_{\textnormal{Se}}^{\textnormal{bulk}}$. Eq. (\ref{Eqformation}) then yields
\begin{equation}
E_f = E(D) - E(0) - n_{\textnormal{Fe}}\mu_{\textnormal{FeSe}} - (n_{\textnormal{Se}}-n_{\textnormal{Fe}})\mu_{\textnormal{Se}},
\end{equation}
where $\mu_{\textnormal{FeSe}} - \mu_{\textnormal{Fe}}^{\textnormal{bulk}} < \mu_{\textnormal{Se}} < \mu_{\textnormal{Se}}^{\textnormal{bulk}}$. Figures~\ref{Fig3}(a), (b) show results for monolayer and bilayer FeSe supercells. In both cases the Fe vacancy possesses the lowest formation energy by a margin of at least $-$0.5 eV within estimated $\mu_{\textnormal{Se}}$ ranges.

\begin{figure}[t]
\includegraphics[scale=1]{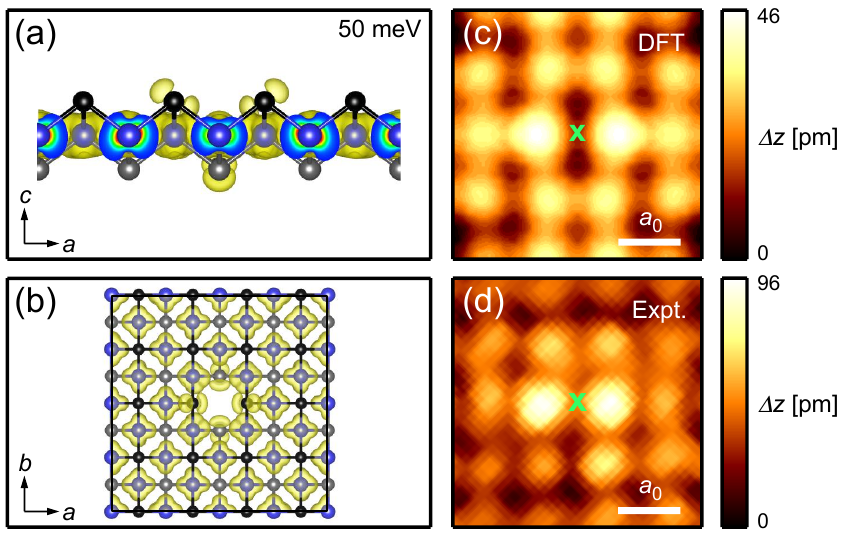}
\caption{(a), (b) Charge density isosurfaces for the Fe vacancy defect configuration, integrated from the Fermi energy up to 50 meV. (c) Simulated STM topography of the Fe vacancy site, marked by a green ``x''; the neighboring Se atoms exhibit brighter lobes, producing a dumbbell signature. (d) Experimental topography (single-layer FeSe/SrTiO$_3$) for comparison. Set point: 50 mV, 500 pA; T = 5 K.}
\label{Fig4}
\end{figure}

%XI. STM topography simulations [Figs. 4(a), (b)]
Having identified the Fe vacancy as the lowest-energy, Se-rich defect of FeSe, we considered whether it can produce a dumbbell signature. Figures~\ref{Fig4}(a), (b) show a charge density isosurface, integrated from the Fermi energy up to 50 meV. For improved accuracy, we increased the BZ sampling to 8$\times$8$\times$1 and used tetrahedron smearing with Bl\"{o}chl corrections~\cite{Blochl_PRB_1994}. Due to the missing Fe atom, orbitals on neighboring Se atoms protrude further out. We simulate an STM topography by tracing the height of the charge density isosurface. As seen in Fig.~\ref{Fig4}(c), the two protruding Se orbitals appear as bright lobes of a dumbbell, matching the experimental image [Fig.~\ref{Fig4}(d)].

%XII. Nudged elastic band setup
If the dumbbell defects are Fe vacancies, they must also be capable of diffusing to the edge of typical film terraces at high temperatures, as implied by Fig.~\ref{Fig1}(e). To elucidate this process, we performed nudged elastic band calculations to find the minimum energy path associated with Fe vacancy hopping~\cite{Mills_SS_1995, Jonsson_book_1998}. We used a smaller $4/\sqrt{2} \times 4/\sqrt{2} \times 1$ supercell with $4 \times 4 \times 1$ BZ sampling. We computed seven intermediate images, each relaxed with total force per atom (tangential and chain) $<$ 0.025 eV/\AA. 

\begin{figure}[b]
\includegraphics[scale=1]{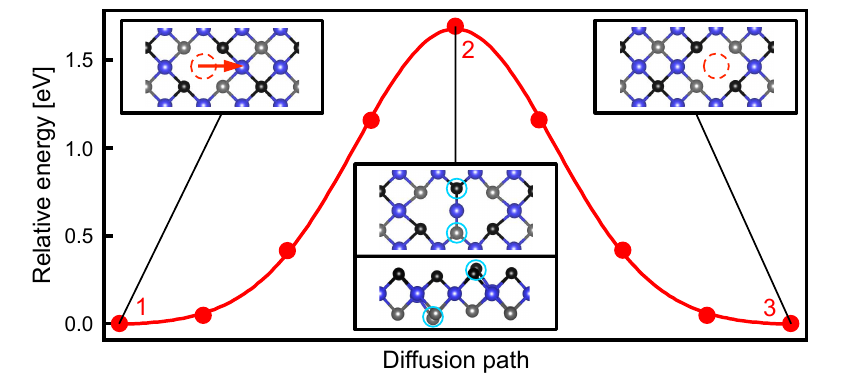}
\caption{Nudged elastic band calculation for nearest-neighbor hopping of an Fe vacancy. The diffusion barrier height is 1.69 eV. Insets depict (1) initial, (2) transition, and (3) final states.}
\label{Fig5}
\end{figure}

%XIII. Fe vacancy diffusion [Fig. 5]
Figure~\ref{Fig5} shows the relative energy along the diffusion path, with insets depicting initial, transition, and final states. In the transition state, two neighboring Se atoms (circled in blue) are pushed above and below the plane, suggesting that vacancy diffusion may be easier on the surface than in the bulk, as expected. We calculate the diffusion rate as
\begin{equation}
\Gamma = \nu \exp\bigg[-\frac{E_B}{k_B T}\bigg],
\end{equation}
where $\nu$ is the attempt frequency and $E_B$ = 1.69 eV is the barrier height. From Vineyard transition-rate theory~\cite{Vineyard_1957, Pandey_PRL_1991} (see Supporting Information),
\begin{equation}
\nu = \bigg(\frac{k_B T}{2 \pi m_{\textrm{Fe}}}\bigg)^{1/2} \bigg[\int_{x_i}^{x_B} dx \exp \bigg[-\frac{E(x)}{k_B T} \bigg] \bigg]^{-1},
\end{equation}
where $m_{\textrm{Fe}}$ is the mass of an Fe atom and $x_i$ ($x_B$) is the initial-state (transition-state) position of the hopping Fe atom. Then for a random walk over a 2D lattice, the root-mean-square distance traveled after time $t$ is
\begin{equation}
x_{\textrm{rms}} = d_{\textrm{Fe-Fe}}\sqrt{\Gamma t},
\end{equation}
where $d_{\textrm{Fe-Fe}} = a/\sqrt{2}$. If we anneal at 450 $^{\circ}$C for 2.5 h [Fig.~\ref{Fig1}(e)], we estimate $x_{\textrm{rms}}$ to be 950 \AA. This distance exceeds typical film island dimensions [Fig.~\ref{Fig1}(c)], thereby explaining how dumbbell defects are removed upon annealing. 

%XIV. Sqrt(5) x Sqrt(5) Ordering
\textbf{Discussion.} We draw a final connection between dumbbell defects and Fe vacancies. At large dumbbell concentrations, Song \textit{et al.}~\cite{Song_PRB_2011} found that the defects ordered into a $\sqrt{5}$$\times$$\sqrt{5}$ superstructure. Similarly, electron diffraction measurements of FeSe crystals, nanosheets, and nanowires have revealed various types of Fe-vacancy order, including $\sqrt{5}$$\times$$\sqrt{5}$$\times$1~\cite{Chen_PNAS_2014}. Given that the Fe vacancy is the thermodynamically most stable defect, the closest packing of these vacancies would lead to a $\sqrt{5}$$\times$$\sqrt{5}$ arrangement, because any closer packing would produce multi-vacancy defects (two or more nearest neighbor Fe atoms mising), which would likely destabilize the crystal altogether. This argument provides an explanation of the $\sqrt{5}$$\times$$\sqrt{5}$ pattern (see Supporting Information for additional calculations). 

%AxFe2-ySe2
The identification of the $\sqrt{5}$$\times$$\sqrt{5}$ dumbbell superstructure with Fe-vacancy order has further significance. Investigations of the related compound $A_x$Fe$_{2-y}$Se$_2$ ($A$ = alkali metal), with enhanced $T_c$ up to 32 K~\cite{Guo_PRB_2010, Wang_PRB_2011}, have been complicated by mesoscale phase separation into multiple Fe-vacancy reconstructions~\cite{Wang_PRB_2011(2), Ricci_PRB_2011, Li_NatPhys_2011, Li_PRL_2012, Ding_NatComm_2013}. Our calculations suggest that Fe vacancy order is not a pathological feature of $A_x$Fe$_{2-y}$Se$_2$, but a phenomenon intrinsic to FeSe grown under excess Se flux. The crucial distinction is that in the latter case, Fe vacancies can be removed upon annealing, while in the former, Fe vacancy diffusion may be hindered by the buffer $A_x$ layers. This additional flexibility in FeSe may afford better control of stoichiometric (superconducting) and ordered vacancy (nonsuperconducting) phases for nanoscale patterning.

%XVI. Summary
In conclusion, we have established the chemical identity of dumbbell defects that appear in MBE-grown FeSe under excess Se flux and suppress superconductivity with concentrations greater than 2.5\%. Our DFT calculations show that Fe vacancies (1) are energetically most favorable, (2) produce dumbbell signatures consistent with STM images, and (3) can diffuse to the edge of typical film islands with vacuum annealing. These atomistic insights lay the foundation towards controlling precise placements of such defects. We also reiterate that amorphous Se may be a poor choice of capping material to perform \textit{ex-situ} measurements due to induced distortions within the underlying FeSe. Finally, we suggest a broader, microscopic connection between dumbbell defect phenomenology in FeSe and mesoscale phase separation in $A_x$Fe$_{2-y}$Se$_2$. 

\begin{acknowledgements}
\textbf{Acknowledgements.} We thank Tetsuo Hanaguri for suggesting to us the Fe vacancy configuration. We thank Shiang Fang, Wei Chen, Matthew Montemore and Joerg Rottler for useful conversations.  This work was supported by the National Science Foundation under Grant No. DMR-1231319 (STC Center for Integrated Quantum Materials), and the Gordon and Betty Moore Foundation's EPiQS Initiative through Grant No. GBMF4536. Computations were run on the Odyssey cluster supported by the FAS Division of Science, Research Computing Group at Harvard University. E. K. acknowledge support by Army Research Office (ARO-MURI) W911NF-14-1-0247. J. E. H. acknowledges support from the Canadian Institute for Advanced Research.
\end{acknowledgements}

\onecolumngrid

\newpage

{\Large \textbf{Supporting Information for:}}
\begin{center}
{\large \textbf{Dumbbell Defects in FeSe Films: \\A Scanning Tunneling Microscopy and First-Principles Investigation}}

Dennis Huang, Tatiana A. Webb, Can-Li Song, Cui-Zu Chang, Jagadeesh S. Moodera, Efthimios Kaxiras, and Jennifer E. Hoffman
\end{center}

\setcounter{figure}{0}
\setcounter{equation}{0}
\setcounter{table}{0}
\makeatletter
\renewcommand{\thefigure}{S\@arabic\c@figure}
\renewcommand{\theequation}{S\@arabic\c@equation}
\renewcommand{\thetable}{S\@arabic\c@table}
\vspace{2mm}

\section{Additional defect configurations}

\begin{figure}[!h]
\includegraphics[scale=1]{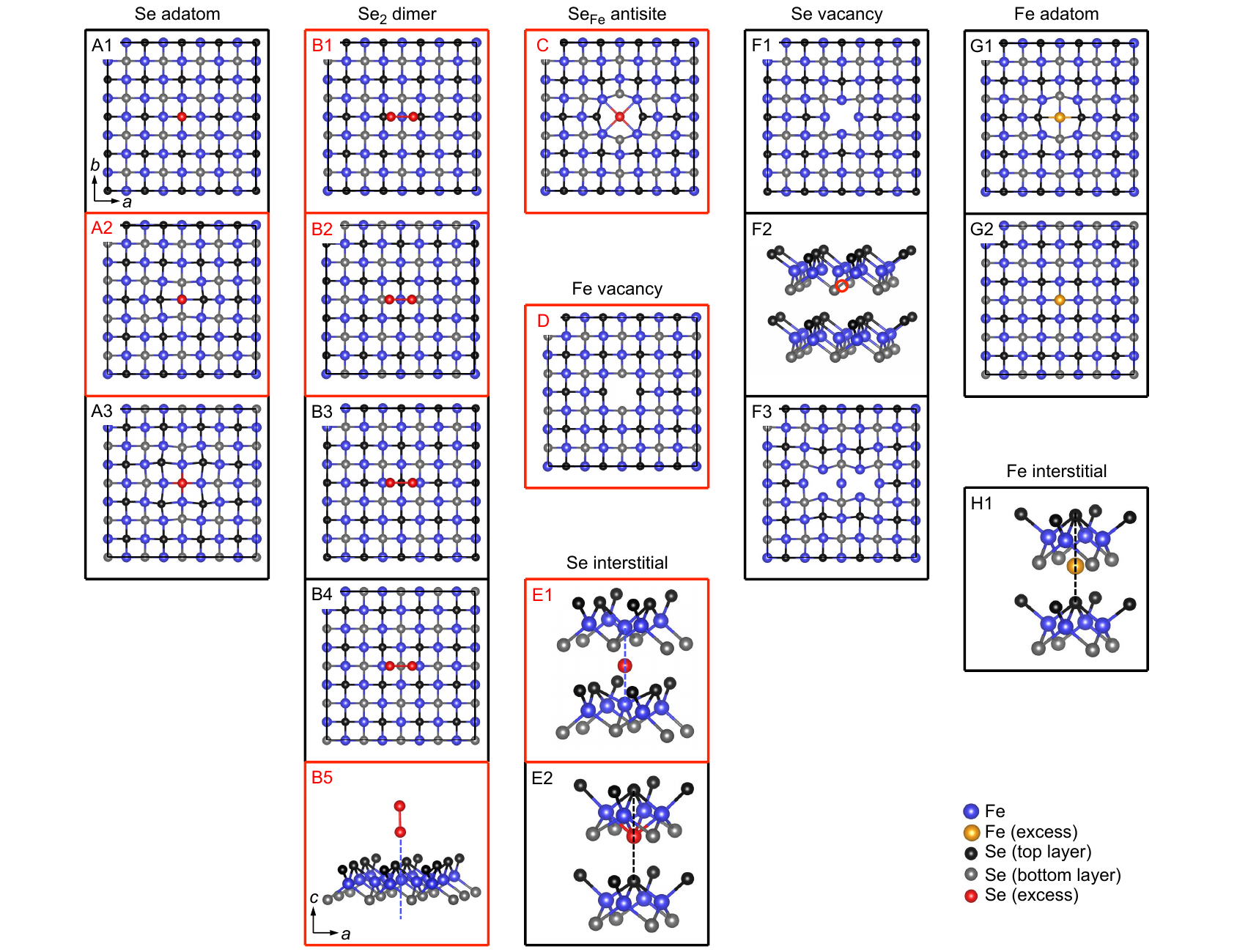}
\caption{Additional relaxed defect structures considered in this work. Panels highlighted in red are Se rich and maintain $C2$-symmetry about the Fe site. Solid-line boxes mark the monolayer FeSe 4$\times$4 supercell boundaries. For (B5), (E1), (E2), (F2), and (H1), only a fraction of the bilayer FeSe 3$\times$3 supercell is shown for clarity. Fe atoms are violet and top/bottom Se atoms are black/gray. Excess Fe/Se atoms are colored orange/red for distinction.}
\label{FigS1}
\end{figure}

\begin{figure}[!h]
\includegraphics[scale=1]{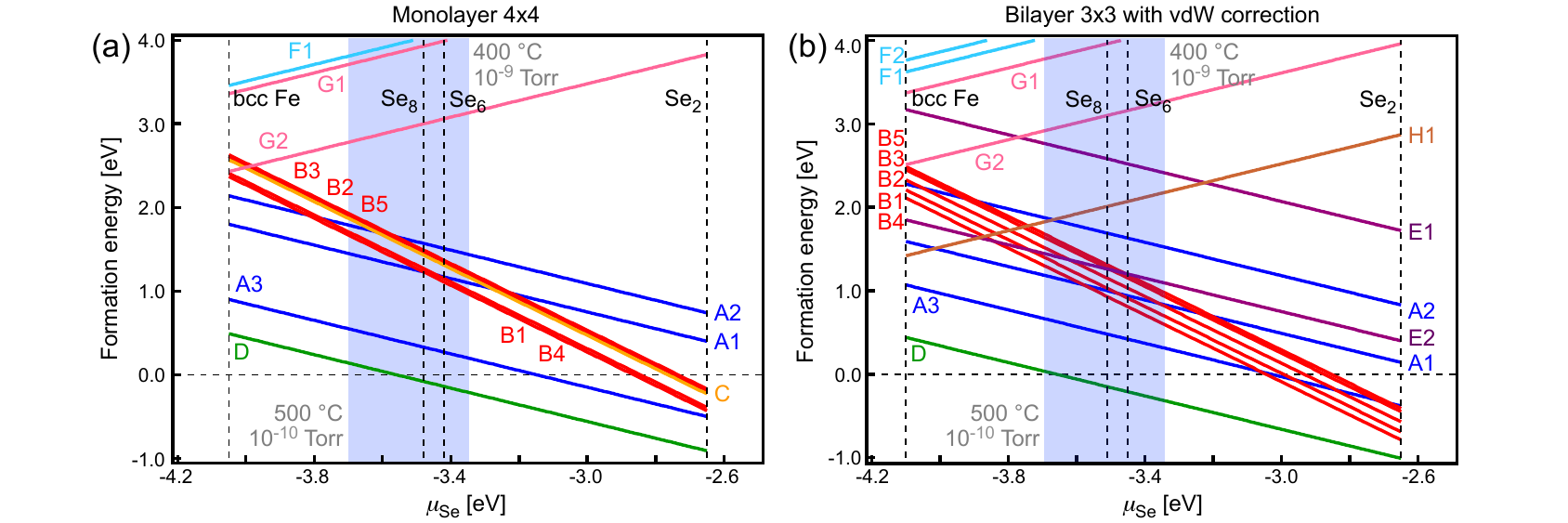}
\caption{Formation energies of defect configurations in the (a) 4$\times$4 supercell monolayer and (b) 3$\times$3 supercell bilayer FeSe. Reproduced from Fig. 3 of the main text, but including additional defect structures from Fig.~\ref{FigS1} whose formation energy is less than 4 eV.}
\label{FigS2}
\end{figure}

Figure~\ref{FigS1} shows all defect configurations examined in this work, which fall into eight categories: (A) Se adatoms, (B) Se$_2$ dimers, (C) Se$_{\textrm{Fe}}$ antisites, (D) Fe vacancies, (E) Se interstitials, (F) Se vacancies, (G) Fe adatoms, and (H) Fe interstitials. Their corresponding formation energies in the monolayer and bilayer FeSe supercells are plotted in Fig.~\ref{FigS2}. Se-rich configurations that are laterally centered at an Fe site, consistent with the STM dumbbell signature, are enclosed in a red box in Fig.~\ref{FigS1}.

\section{Selenium chemical potential}

%Formalism
Following Ref.~\cite{Reuter_PRB_2001}, we estimate the experimental chemical potential of Se using ideal gas approximations and tabulated thermodynamic quantities. Under fixed temperature $T$ and pressure $p$, the chemical potential of an ideal gas of Se$_2$ molecules is related to the Gibbs free energy as
\begin{equation}
\mu_{\textrm{Se}}(T, p) = \frac{G_{\textrm{Se}_{2(g)}}(T, p)}{2N},
\end{equation}
where $N$ is the number of Se$_2$ molecules. First, we compute $G_{\textrm{Se}_{2(g)}}$ at standard conditions ($T^{\circ}$ = 298.15 K, $p^{\circ}$ =  1 bar),
\begin{equation}
G_{\textrm{Se}_{2(g)}} (T^{\circ}, p^{\circ}) = H_{\textrm{Se}_{2(g)}} (T^{\circ}, p^{\circ}) - T^{\circ} S_{\textrm{Se}_{2(g)}} (T^{\circ}, p^{\circ}),
\end{equation}
using enthalpy $H_{\textrm{Se}_{2(g)}} (T^{\circ}, p^{\circ})$ and entropy $S_{\textrm{Se}_{2(g)}} (T^{\circ}, p^{\circ})$ values derived from thermochemistry references. Next, we extrapolate to desired temperatures using reference heat capacity data, 
\begin{equation}
G_{\textrm{Se}_{2(g)}} (T, p^{\circ}) = H_{\textrm{Se}_{2(g)}} (T^{\circ}, p^{\circ}) + \int_{T^{\circ}}^{T} dT' C_{p^{\circ}, \textrm{Se}_{2(g)}}(T') - T \bigg[ S_{\textrm{Se}_{2(g)}}(T^{\circ}, p^{\circ}) + \int_{T^{\circ}}^{T} dT' \frac{C_{p^{\circ}, \textrm{Se}_{2(g)}}(T')}{T'} \bigg],
\end{equation}
and desired pressures using ideal gas relationships,
\begin{equation}
G_{\textrm{Se}_{2(g)}} (T, p)  = G_{\textrm{Se}_{2(g)}} (T, p^{\circ}) + N k_B T \textrm{ln}\bigg(\frac{p}{p^{\circ}}\bigg).
\end{equation}

%Practical computation notes.
We make some practical remarks on estimating $\mu_{\textrm{Se}}(T, p)$:

(1) To maintain consistency with DFT calculations and total energies ($E$) defined in VASP, we set $\mu_{\textrm{Se}}(0~\textrm{K}, p) = E_{\textrm{Se}}= 0$ for an isolated Se \textit{atom}.

(2) Based on the chosen reference, we compute $H_{\textrm{Se}_{2(g)}} (T^{\circ}, p^{\circ})$ in a two-step process:
\begin{equation}
\begin{cases}
       2\textrm{Se}_{(g)}(0~\textrm{K}, p^{\circ}) \rightarrow 2\textrm{Se}_{(g)}(T^{\circ}, p^{\circ}), & \Delta H_1 = 2(\frac{5}{2} k_B T^{\circ}),~\textrm{ideal monoatomic gas}, \\
       2\textrm{Se}_{(g)}(T^{\circ}, p^{\circ}) \rightarrow \textrm{Se}_{2(g)}(T^{\circ}, p^{\circ}), & \Delta H_2 = 2(-165.520 \pm 0.250)~\textrm{kJ} \textrm{mol}^{-1},~\textrm{p. 53 of Ref.~\cite{Olin_2005}}.
\end{cases}
\end{equation}

(3) From p. 40 of Ref.~\cite{Olin_2005}, $S_{\textrm{Se}_{2(g)}} (T^{\circ}, p^{\circ})$ = $247.380 \pm 0.400$ JK$^{-1}$mol$^{-1}$.

(4) From p. 63 of Ref.~\cite{Olin_2005}, $C_{p^{\circ}, \textrm{Se}_{2(g)}}(T) = a + bT + cT^2 + dT^{-1} + eT^{-2}$, where $a$ = $1.93485 \times 10$ JK$^{-1}$mol$^{-1}$, $b$ = $1.24903 \times 10^{-2}$ JK$^{-2}$mol$^{-1}$, $c$ = $-2.07010 \times 10^{-6}$ JK$^{-3}$mol$^{-1}$, $d$ = $1.09846 \times 10^{4}$ Jmol$^{-1}$, and $e$ = $-1.60249 \times 10^{6}$ JKmol$^{-1}$. These values are valid from $T$ = 298 K to 1300 K.

(5) We choose $T$ to be the substrate temperature and $p$ to be the Se partial pressure in the chamber, based on quasi-equilibrium growth assumptions suggested for similar III-V semiconductor MBE processes~\cite{Kratzer_PRB_1999, Tersoff_PRL_1997}. For $T$ = 400 $^{\circ}$C and $p$ = 10$^{-9}$ T, we find that $\mu_{\textrm{Se}}$ = $-$3.35 eV. For $T$ = 500 $^{\circ}$C and $p$ = 10$^{-10}$ T, we find that $\mu_{\textrm{Se}}$ = $-$3.70 eV.

\section{Simulated STM topographies}

\begin{figure}[!h]
\includegraphics[scale=1]{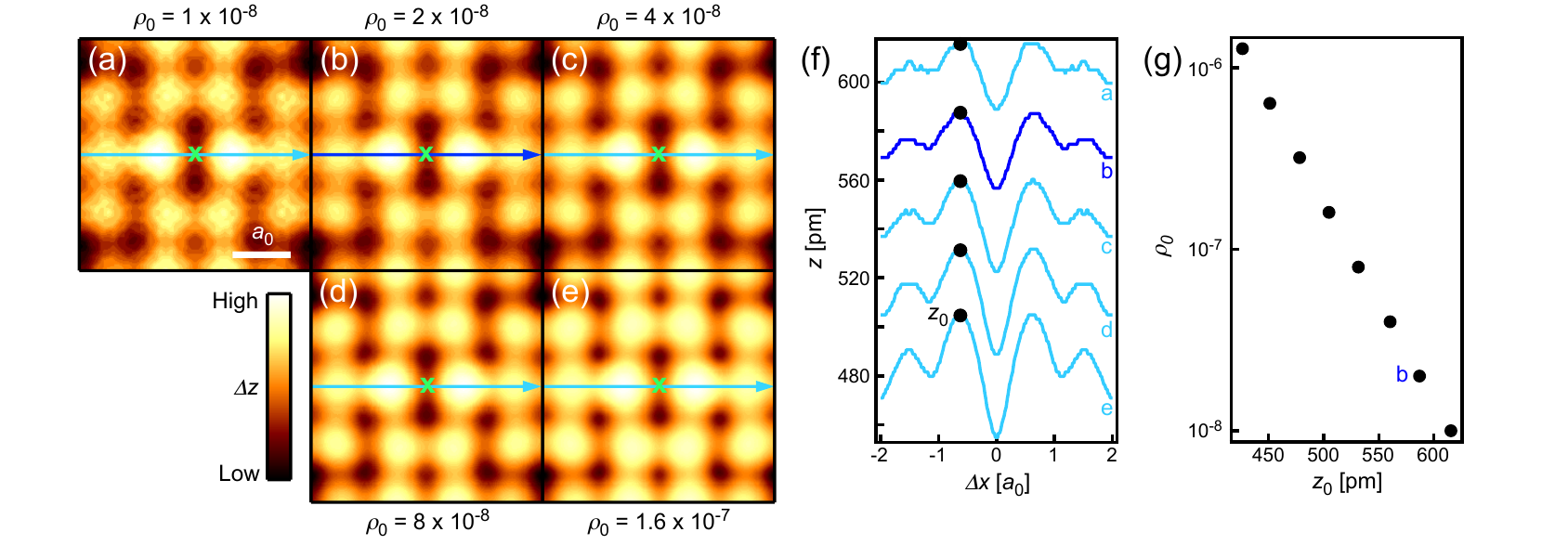}
\caption{(a)-(e) Simulated STM topographies of an Fe vacancy site, with different isosurfaces of charge density ($\rho_0$). (f) Horizontal line cuts across the defect center. $z$ is the distance measured from the Fe plane. The dark blue contour (b) depicts the isosurface chosen in Fig. 4 of the manuscript. (g) Plot of $\rho_0$ vs. the maximum height ($z_0$) of the dumbbell impurity, demonstrating that our simulations are in the regime of exponential decay.}
\label{FigS3}
\end{figure}

To simulate the STM topography of an Fe vacancy site, we use DFT to compute the real-space charge density, integrated from the Fermi energy up to 50 meV. We then trace the height variation ($\Delta z$) for a given isosurface of charge density ($\rho_0$). As shown in Fig.~\ref{FigS3}, the value of $\rho_0$ chosen does not qualitatively affect the simulated topography, but we pick $\rho_0$ to be clearly in the regime of exponential decay.    

\begin{figure}[!h]
\includegraphics[scale=1]{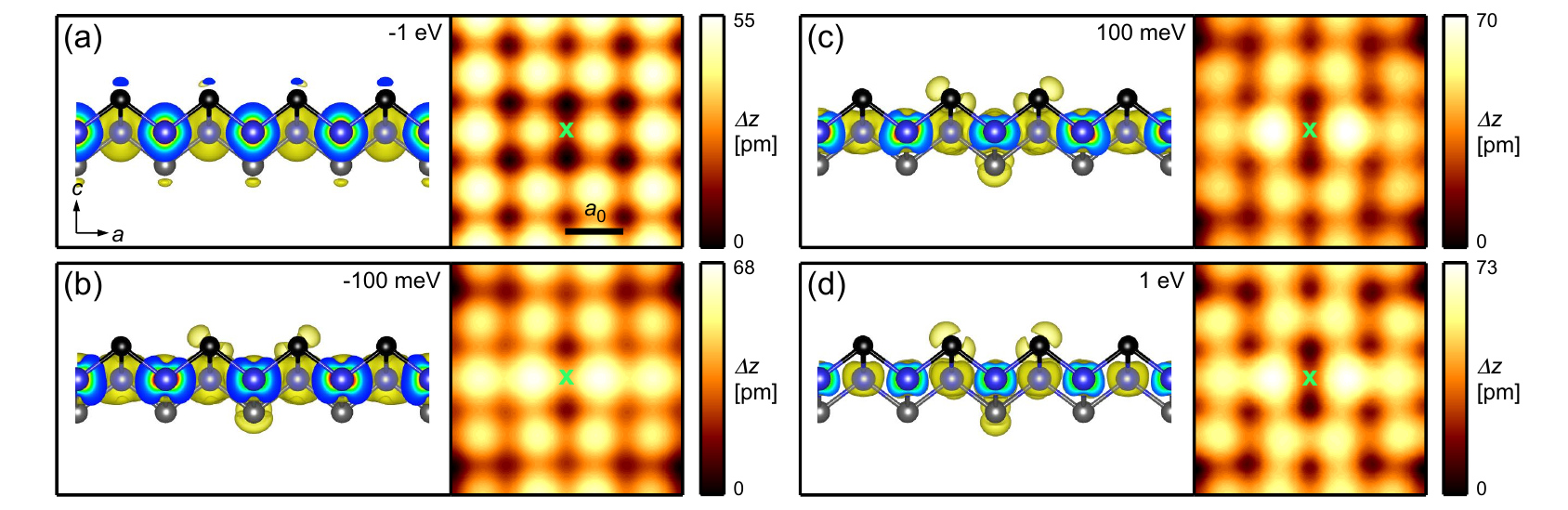}
\caption{Charge density isosurfaces and simulated STM topographies of an Fe vacancy, calculated for different imaging biases. The Fe vacancy site is marked by a green ``x.''}
\label{FigS4}
\end{figure}

Figure~\ref{FigS4} shows STM topographic simulations of an Fe vacancy with different imaging biases $V$, carried out by integrating the charge density from the Fermi energy to $eV$. We note that experimental FeSe bands exhibits orbital-dependent renormalization with an average factor $1/z$ = 1/6 relative to LDA/GGA-calculated bands~\cite{Maletz_PRB_2014, Watson_PRB_2015, Mukherjee_PRL_2015}, precluding a more detailed \textit{ab-initio} analysis of STM defect signatures.

\section{Vineyard transition-rate theory}

The classical probability distribution for an ensemble of identical, non-interacting particles moving in a 1D potential $\Phi(x)$ is given by 
\begin{equation}
\label{Eqdistr}
p(x, v) = \rho_0 \bigg(\frac{m}{2\pi k_B T}\bigg)^{1/2} \exp \bigg[ -\frac{\Phi(x)}{k_B T} - \frac{mv^2}{2 k_B T}\bigg],
\end{equation}
where $m$ is the particle mass, $v$ is its velocity, and $\rho_0$ is a normalization constant. We consider a finite spatial region in phase space, $x \in [x_i, x_f]$, with a potential barrier peaked at $x_B \in (x_i, x_f)$. The average transition rate $\Gamma_{x_i \rightarrow x_f}$ is given by $I/Q$, where $I$ is the phase space current across the barrier and $Q$ is the number of states in $[x_i, x_B]$~\cite{Vineyard_1957}. From Eq. (\ref{Eqdistr}), we find that
\begin{equation}
Q =  \int_{x_i}^{x_B} dx \int_{-\infty}^{\infty} dv p(x,v) = \rho_0  \int_{x_i}^{x_B} dx \exp \bigg[-\frac{\Phi(x)}{k_B T} \bigg] 
\end{equation}
and
\begin{equation}
I = \int_{0}^{\infty} dv v p(x_B,v) = \rho_0 \exp \bigg[-\frac{\Phi(x_B)}{k_B T} \bigg] \bigg(\frac{k_B T}{2 \pi m}\bigg)^{1/2},
\end{equation}
which yields
\begin{equation}
\Gamma_{x_i \rightarrow x_f} = \bigg(\frac{k_B T}{2 \pi m}\bigg)^{1/2} \bigg[\int_{x_i}^{x_B} dx \exp \bigg[-\frac{\Phi(x)}{k_B T} \bigg] \bigg]^{-1} \exp \bigg[-\frac{\Phi(x_B)}{k_B T}\bigg].
\end{equation}
In our application to an Fe atom hopping to a neighboring vacant site in FeSe, we take $m = m_{\textrm{Fe}}$, and $\Phi(x)$ to be the DFT total energy $E$ along the diffusion path, parameterized by the position $x$ of the moving Fe atom~\cite{Kaxiras_PRB_1993}.

\section{$\sqrt{5}$$\times$$\sqrt{5}$ vacancy order}

\begin{figure}[!h]
\includegraphics[scale=1]{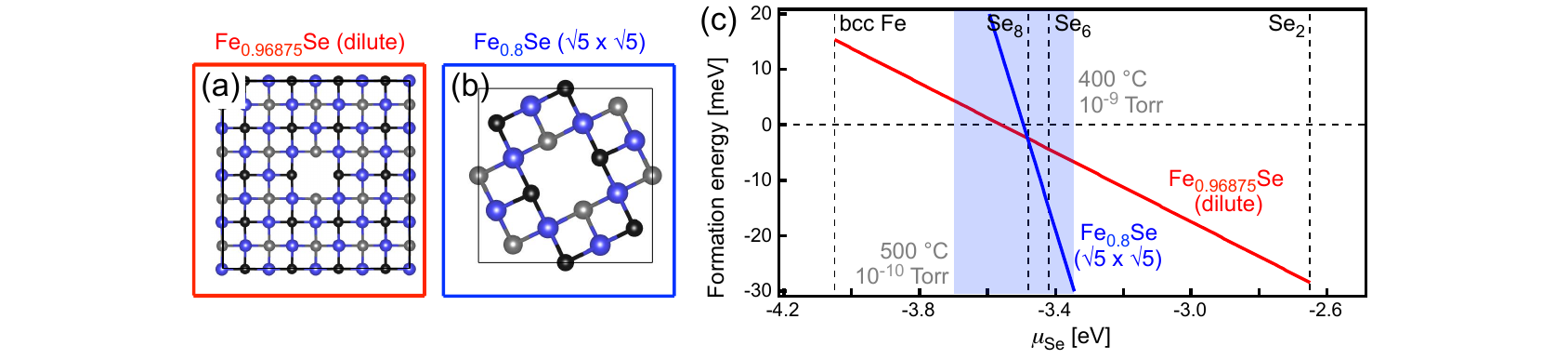}
\caption{Formation energies of Fe$_{1-x}$Se$_x$ for two concentrations of Fe vacancies: (a) $x$ = 0.03125 (dilute), and (b) $x$ = 0.2 ($\sqrt{5}$$\times$$\sqrt{5}$ order). (c) Assuming no condensation of bulk Fe (body-centered cubic) or Se (Se$_6$ and Se$_8$ rings), chemical potential values $\mu_{\textrm{Se}}$ are restricted between the dashed lines labeled bcc Fe and Se$_6$/Se$_8$. $\mu_{\textrm{Se}}$ is set to zero for an isolated Se atom. Alternatively, shaded blue regions mark estimated $\mu_{\textrm{Se}}$ values at typical substrate temperatures and Se partial pressures, using ideal gas approximations and tabulated thermodynamic quantities.}
\label{FigS5}
\end{figure}

We show that a $\sqrt{5}$$\times$$\sqrt{5}$ ordering of Fe vacancies in FeSe is thermodynamically stable under excess Se flux. Following the same reasoning in the main text, the formation energy of Fe$_{1-x}$Se, where $0 \le x \le 1$, is given by
\begin{equation}
E_f = E(\textnormal{Fe$_{1-x}$Se}) - E(\textrm{FeSe}) + x(\mu_{\textnormal{FeSe}} - \mu_{\textnormal{Se}}).
\end{equation}
We consider two concentrations:

(1) Dilute: $x$ = 0.03125. This value corresponds to one Fe vacancy in a monolayer 4$\times$4 supercell [Fig.~\ref{FigS5}(a)]. 

(2) $\sqrt{5}$$\times$$\sqrt{5}$ Fe vacancy order: $x$ = 0.2. For this DFT calculation, we use a monolayer $\sqrt{5}$$\times$$\sqrt{5}$ supercell [Fig.~\ref{FigS5}(b)] with 5$\times$5$\times$1 BZ sampling. 

Figure~\ref{FigS5}(c) demonstrates that within a narrow range of viable $\mu_{\textnormal{Se}}$ values, the $\sqrt{5}$$\times$$\sqrt{5}$ Fe vacancy superstructure is more energetically favorable compared to a dilute Fe vacancy.

\end{document}